\documentstyle[twocolumn,prb,aps,epsf]{revtex}

\begin{document}
%********************************************************************
\title{\bf
\vspace*{-10mm}
\begin{flushright}
{\large {\bf PREPRINT:} 
Applied Physics Report 98-40; Submitted to Phys. Rev. B}
\end{flushright}
\vspace*{5mm}
Far-infrared induced current
in a ballistic channel -- potential barrier structure.
}
\author{Ola Tageman, and A. P. Singh
\\  Department of Applied Physics, Chalmers University of Technology
\\  and G\"oteborg University, S-41296 G\"oteborg, Sweden }

%\begin{document}
\maketitle

\begin{abstract}
We consider electron transport in a ballistic multi-mode channel
structure in the presence of a transversely 
polarized far-infrared (FIR) field.
The channel structure consists of a long resonance region
connected to an adiabatic widening with a potential barrier
at the end.
At frequencies that match the mode energy separation
in the resonance region we find distinct 
peaks in the photocurrent, caused by
Rabi oscillations in the mode population.
For an experimental situation in which the
width of the channel is tunable via gates, we propose a method
for reconstructing the spectrum of propagating modes,
without having to use a tunable FIR source.
With this method the change in the spectrum as the
gate voltage is varied can be monitored.

\end{abstract}

%********************************************************************
\section{Introduction}
%********************************************************************

Attempts have been made during the past few years
to observe effects of the application of a far-infrared (FIR) field
on electrons propagating coherently in narrow quantum channels.
\cite{Wyss:APL:93,Janssen:JPC:94,Wyss:APL:95,Arnone:APL:95}
In focus of such investigations has been
quantum point contacts (QPCs)
defined by split-gate depletion of a
two-dimensional electron gas in a GaAs/AlGaAs heterointerface.
\cite{Wees:PRL:88,Wharam:JPC:88}
So far the experiments employing both transverse and parallel
polarization have been explained by heating and
rectification effects, while
clear evidence of FIR-field influence 
inside the channel is still missing.

Considering the pronounced transverse quantization
into waveguide modes it becomes particularly interesting to
consider transverse polarization of the FIR field, since it
allows strong mode coupling even for a 
homogeneous FIR field.
For transverse polarization it is well known from 
theoretical considerations that
transitions between modes take place mainly at points
where the mode energy separation, $E_m-E_n$,
equals the photon energy $\hbar\omega$.
\cite{Hekking:PRB:91,Gorelik:PRL:94,Grincwajg:PRB:95}
At such points momentum conservation is possible.
In the case of a QPC the mode energy separation changes
and resonance conditions can be fulfilled only in a small
region, which reduces the sensitivity to the FIR field.
This could explain why not even experiments using transverse
polarization have succeeded in providing evidence of
coherent FIR-field influenced transport.

The natural way to increase the sensitivity to a FIR field
is to extend the region of resonance by considering a longer 
channel in which the width is constant.
To consider long coherent channels has in  the last few
years become realistic due to progress in fabrication techniques.
\cite{Tarucha:JJAP:95}
Since FIR-field induced transitions between propagating modes alone
will not change the transmission probability, it is
necessary to incorporate some detection mechanism that discriminates
between modes. This idea has been developed in the special
case when only one mode enters the channel and a QPC is
used for detection, which leads to a reflection of 
excited electrons.
\cite{Tageman:JAP:97,Tageman:JAP:98}

In this work we consider the complementary case of a
multi-mode channel.
A new feature is that there is a possibility for 
cancellation if one electron climbs the mode spectrum while
another one descends.
Since there is no lower limit to the mode energy separation
when the restriction of only one propagating mode is relaxed,
a wider range of frequencies is of interest.
For detection we use a combination of an adiabatic widening
and a line-gate barrier, which leads to an enhanced transmission
for excited electrons.
We find a general expression for the scattering states,
that includes all relevant mode mixing, 
by solving an eigenvalue problem.
An explicit analytical expression is derived which is 
valid in the limit of weak
FIR-fields, where resonances between pairs of 
transverse modes will dominate.
We also propose an experimental 
method for finding the mode spectrum
using only fixed frequency FIR-sources.

%********************************************************************
\section{Model system}
\begin{figure}[htb] \begin{center}\leavevmode
\epsfxsize 0.85 \hsize
\epsfbox{
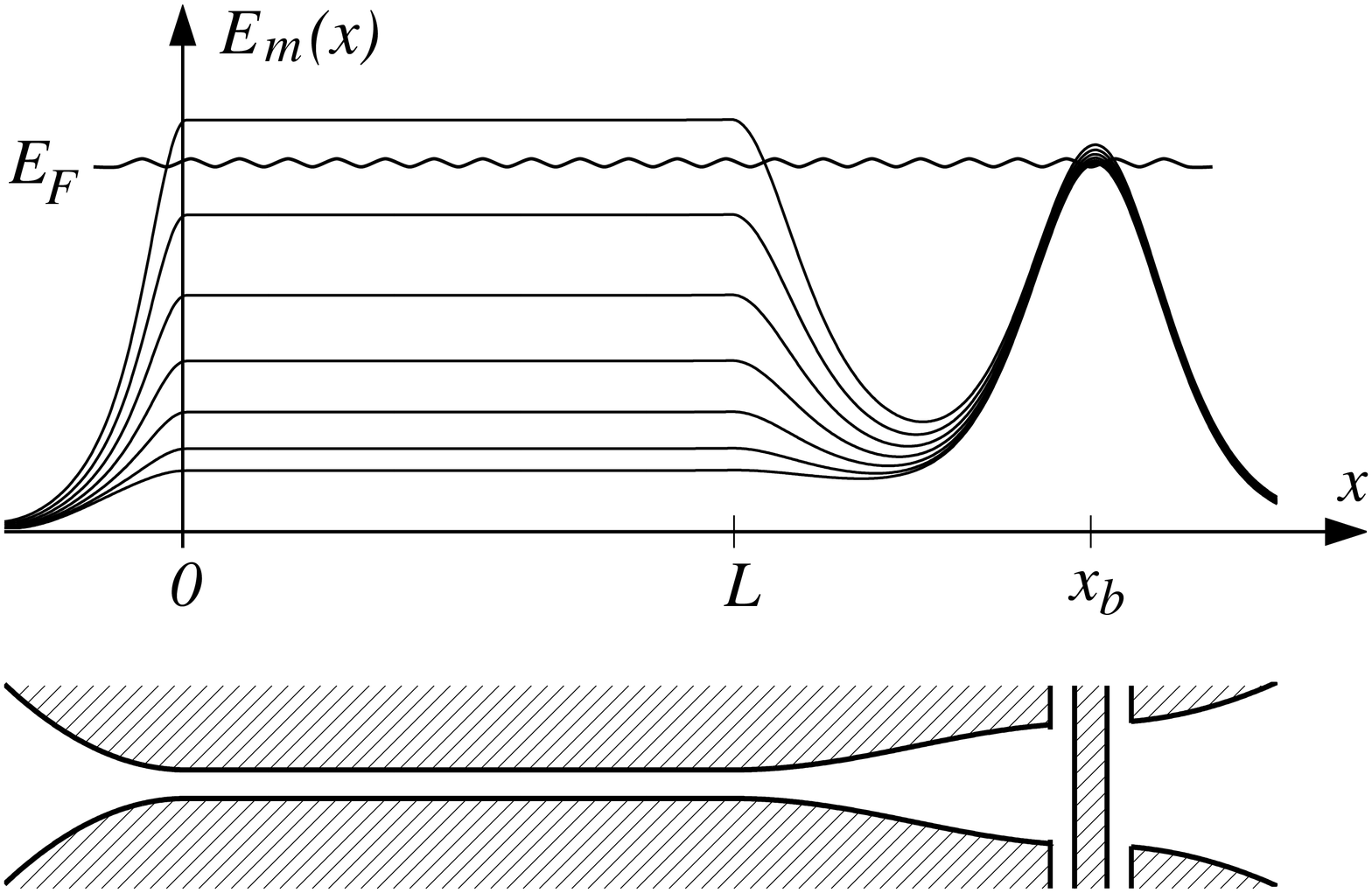
}
\vspace{2ex}
\caption{
\label{fig:denergies}
Geometry of the channel and mode energies $E_m(x)$.
Between $x=0$ and $x=L$ there is a resonance region with
an $x$-independent mode energy spacing.
At $x_b$ there is a line-gate which gives rise to the
barrier in the potentials.
}
\end{center}
\end{figure}

%********************************************************************
We consider a quasi-onedimensional channel 
that smoothly connects two reservoirs.
The channel consist of three regions, each with its own purpose
(see Fig. \ref{fig:denergies}).
First, there is a resonance region featuring strong
size quantization, in which an external 
high frequency electric field polarized
across the channel can induce mode transitions.
Second, there is an adiabatic widening in which there
is a conversion of the energy stored in the transverse direction
into kinetic energy for longitudinal motion.
Finally, there is a barrier that blocks slow electrons.

Experimentally such a system can be realized in different ways,
but in order to see resonances fully developed it takes a system
that preserves coherence all along the channel. We primarily 
have a split-gate channel in a GaAs/AlGaAs-interface in mind.
The barrier can then be created by a line gate across the channel.
Unlike a split-gate barrier it gives the same barrier for all modes.
The height of this
barrier is adjusted so that no transport takes
place in the absence of the FIR-field. Then only those electrons
that have been excited can pass. Since there is mode pumping
on one side only a photocurrent will be generated even in absence
of a driving voltage. We are interested in this zero bias 
photocurrent.

We assume that the width of the channel varies slowly on the
scale of the electron wavelength which allows us to separate the
transverse and the longitudinal motion.
\cite{Glazman:JETP:88,Glazman:PRB:90}
Phase breaking processes and collective effects are ignored,
and the current is calculated from single particle 
transmission probabilities in the spirit of the Landauer approach.
\cite{Landauer:PS:92}

%********************************************************************
\section{Multi-mode Scattering states}
%********************************************************************
We solve a single particle Schr\"odinger equation with the following
time dependent Hamiltonian:
\begin{equation}
\hat{H}(t)=
\frac{1}{2m^*}(\hat{{\bf p}}-\frac{e}{c}{\bf A}(x,t))^2+U(x,y)
\;.
\label{eq:Hamiltonian}
\end{equation}
Here
${\bf \hat{p}}$ is the momentum,
$m^*$ is the effective mass
of an electron in the two-dimensional electron gas
and the potential $U(x,y)$
confines the electron to the channel
and reservoirs.
The vector potential $ {\bf A}(x,t)$ in Eq. (\ref{eq:Hamiltonian})
describes a homogeneous electromagnetic
field of angular frequency $\omega$ and amplitude $\hat{E}$
polarized in the y-direction:
\begin{equation}
{\bf A}(x,t)=\frac{ c}{\omega}\, \hat{E} 
\cos(\omega t)\, {\bf e}_y \;.
\end{equation}

We now make a separation Ansatz for the wave function:
\begin{equation}
\Psi(x,y,t)=\sum_m \Psi_m(x,t)\Phi_m(x,y)\;.
\label{eq:separation}
\end{equation}
Here $\Phi_{m}(x,y)$ are solutions to the
transverse eigenvalue equation:
\begin{eqnarray}
-\frac{\hbar^2}{2m^*}\frac{\partial^2}{\partial y^2}
\Phi_m(x,y)
+
U(x,y)\Phi_m(x,y)
\nonumber \\
=
E_m(x)\Phi_m(x,y)\;.
\end{eqnarray}
If the channel geometry changes slowly along the channel we
can neglect $x$-derivatives of $\Phi_m(x,y)$ when inserting
the Ansatz into the Schr\"odinger equation and find:
\cite{Glazman:JETP:88}
%({\em eq:longitudinal})I
%
\begin{eqnarray}
-\frac{\hbar^2}{2m^*}
\frac{\partial^2}{\partial x^2}\Psi_m(x,t)
+E_m(x)\Psi_m(x,t)
-i\hbar\frac{\partial}{\partial t}\Psi_m(x,t)
\nonumber \\
=
-2i\cos(\omega t)\sum_{m}M_{mm'}(x)\Psi_{m'}(x,t)\;,
\label{eq:longitudinal}
\end{eqnarray}
where the intermode transition elements $M_{mm'}(x)=-M_{m'm}(x)$
are given by:
%({\em eq:Matrixelement})
%
\begin{equation}
M_{mm'}(x)=\frac{\hbar e {\hat{E}}}{2 m^*\omega}
\int \Phi_m^*(x,y) \frac{\partial}{\partial y} 
\Phi_{m'}(x,y) dy\;.
\label{eq:Matrixelement}
\end{equation}
%

%********************************************************************
%\subsection*{Resonance region}
%********************************************************************
We now look for solutions to Eq. (\ref{eq:longitudinal})
in the resonance region where $U(x,y)$ is assumed to be 
$x$-independent and use the following Ansatz:
\begin{equation}
\Psi_{m}(x,t)
=
\sum_l
C_{ml}\, e^{i\left[ Px-(E+l\hbar\omega)\; t 
\right] /\hbar}
\;.
\label{eq:solution}
\end{equation}
When this Ansatz is inserted into Eq. (\ref{eq:longitudinal})
we get an infinite set of coupled equations for the coefficients
$C_{ml}$:
%
%({\em eq:sys})
%
\begin{eqnarray}
\lefteqn{
\left(
P^2/2m^*-K_{ml}(E)
\right)
C_{ml}
}\hspace{10mm}&&
\label{eq:sys} \\
&&=
-\sum_m iM_{mm'}
\left(
C_{m'(l-1)}+C_{m'(l+1)}
\right)\;,
\nonumber
\end{eqnarray}
where:
%
%({\em eq:kinenergy})
\begin{equation}
K_{ml}(E)
=
E+l\hbar \omega-E_m\;,
\label{eq:kinenergy}
\end{equation}
Eq. (\ref{eq:sys})
can be put in the form of an eigenvalue problem:
%
%({\em eq:matek})
%
\begin{equation}
\hat{K}_{ml,m'l'}(E)\, C_{m'l'}=P^2/2m^* C_{ml}
\;,
\label{eq:matek}
\end{equation}
where:
%
%({\em eq:matten})
%
\begin{eqnarray}
\hat{K}_{ml,m'l'}(E)&=&K_{ml}(E)\,\delta_{mm'}\delta_{ll'}
\nonumber \\
&&-iM_{mm'}(\delta_{l(l'+1)}+\delta_{l(l'-1)})
\;,
\label{eq:matten}
\end{eqnarray}
with $\delta$ being the Kronecker delta.

We will now use the fact that the resonance region
has a finite length $L$, and that our solution inside the resonance
region arise due to an incoming wave from the reservoir of
the following form:
%
%({\em eq:incoming})
%
\begin{eqnarray}
\Psi_{n,E}^{INC}(x,y,t)=
\Phi_n(x,y)
\frac{
e^{i\left[ \int p_n(x) dx -Et
\right]/\hbar}
}{\sqrt{p_{n}(x)/p_{n}(0)}
}
\;,
\label{eq:incoming}
\end{eqnarray}
where $p_n(x)=\sqrt{2m^*(E-E_n(x))}$.
This in combination with the restriction: $|M|\ll \hbar\omega$,
allows us to get a good approximation
to the scattering state using 
a reduced number of states in the eigenvalue problem in 
Eq. (\ref{eq:matek}).

We pay attention to resonance effects only.
For this reason we 
keep only one coefficient $C_{ml}$ for each mode, i.e.
the one that gives the best kinetic energy match with
the incoming state.
Explicitly we define $l_{mn}$ as the
integer that minimizes $|E_m-E_n+l_{mn}\hbar\omega|$, and
keep only coefficients $C_{ml_{mn}}$.

By this approach
we miss the possibility of
indirect resonant coupling via non-resonant states.
On the other hand all ignored states necessarily give
a kinetic energy mismatch of at least $\hbar\omega/2$.
This leads to a reduction of the 
effective coupling, via one or several such
states, by at least a factor of the order 
$|M|/\hbar\omega$. 
The characteristic length scale for population oscillation is
at resonance:
$\lambda_R=\lambda \cdot K/|M|$, where $\lambda$ is the de Broglie wavelength
and $K$ is the kinetic energy.
\cite{Tageman:JAP:97}
Obviously this length scale is a 
factor of $\hbar\omega/|M|$ larger for the indirect resonant
coupling than for the direct one.
Therefore there is a regime
of weak FIR fields in which $\lambda_R^{IND}\gg L$,
where it suffices to keep only the
``most resonant'' coefficients.

We consider a situation in which 
the transport is adiabatic outside the resonance
region. This is realistic if the FIR field is sufficiently weak.
Far from the resonance region, FIR field 
induced transitions are suppressed
by a factor of the order $|M| /\hbar\omega$, 
because of a kinetic mismatch of the order $\hbar\omega$.
Near the resonance region the mode potentials
are taken to be
rapidly varying on the scale of $\lambda_R$ (but still slowly on
the scale of $\lambda$).
In this case there is no room, for a resonance
to develop a significant change in 
the mode population.
One can think of the FIR field as
being suddenly switched on at $x=0$ and suddenly switched
off at $x=L$.
In particular we can assume
that the incident mode, $n$, is still fully
occupied at $x=0$ (where the resonance region begins)
and that the mode population at $x=L$ (where the 
resonance region ends) will remain for $x>L$.

In the resonance region we get,
using Eq. (\ref{eq:separation}),
the following expression for the 
scattering wave function:
%({\em eq:scattsol})
%
\begin{eqnarray}
\Psi_{n,E}^{RES}(x,y,t)=
\sum_{rm} a_{n,E}^{(r)}\Phi_m(y)
C_{n,E,m}^{(r)} 
\nonumber
\\
\times
e^{i\left[ P_{n,E}^{(r)} x-(E+\hbar\omega l_{nm})t
\right]/\hbar}
\;,
\label{eq:scattsol}
\end{eqnarray}
where $P_{n,E}^{(r)}=\sqrt{2m^*{\cal K}_{n,E}^{(r)}}\,$, and
${\cal K}_{n,E}^{(r)}$ and 
$C_{n,E,m}^{(r)}$ are the 
eigenvalues and the orthonormal eigenvectors
of the following reduced eigenvalue equation:
%({\em eq:smalleigsys})
%
\begin{equation}
\sum_{m}\left[
K_{ml_{mn}}(E)\,\delta_{mm'}
-M_{mm'}
\right] \,
C_{m'}={\cal K}\, C_{m}
\;,
\label{eq:smalleigsys}
\end{equation}
and the coefficients $a_{n,E}^{(r)}$
are chosen such that only mode $n$ is populated at $x=0$:
%({\em eq:match})
%
\begin{eqnarray}
\delta_{mn}
=
\sum_{r} a_{n,E}^{(r)}
C_{n,E,m}^{(r)} 
\;.
\label{eq:match}
\end{eqnarray}

In the widening region ($x>L$) we get, using the assumption
of adiabaticity,
the following expression for
the scattering wave function:
%({\em eq:after})
%
\begin{eqnarray}
\lefteqn{
\Psi_{n,E}^{WID}(x,y,t)=\sum_{m} t_{mn}(E)
}\hspace{5mm}&&
\nonumber \\
&&
\times
\Phi_m(x,y)
\frac{
e^{i\left[ \int p_{n,E,m}(x)dx-(E+\hbar\omega l_{mn})t
\right]/\hbar}
}{\sqrt{p_{n,E,m}(x)/p_{n,E,m}(L)}
}
%\;,
\label{eq:after}
\end{eqnarray}
where $p_{n,E,m}(x)=\sqrt{2m^*(E+\hbar\omega l_{mn}-E_m(x))}$, and
the transition amplitudes $t_{mn}$ are given by:
%
%({\em eq:scattmatel})
%
\begin{eqnarray}
t_{mn}(E)
=
\sum_{r} a_{n,E}^{(r)}
C_{n,E,m}^{(r)} e^{i P_{n,E}^{(r)} L
/\hbar}
\;.
\label{eq:scattmatel}
\end{eqnarray}

When we call $t_{mn}(E)$ transition amplitudes we
implicitly assume that scattering takes place between states
with the same momentum.
The coupling is strong
only between states whose difference in kinetic energy $\Delta K$ 
fulfills
$\Delta K < |M|$. Thus for coupled states we have
$\Delta p/p\approx \Delta K/2K<|M|/2K$, and assuming $|M|\ll K $
we can forget this difference in momentum.

Our scattering solution $\Psi_{n,E}(x,y,t)$
is valid for sufficiently weak FIR fields since it does not
take reflections into account.
In general there will be reflections both at the entrance 
and at the exit, but they will be suppressed if $|M|\ll K $.
In principle we could
handle more than the weak FIR field limit by
using an exact matching condition that
takes into account also the derivative of the wave function,
However, in the strong field limit
one would also have to account for
multiple reflection from the entrance, the exit and
from the barrier.

An electron near the band bottom will not fulfill
the condition $|M|\ll K $. For such an electron
we fail to make reliable predictions.
On the other hand,
since each energy interval is weighed equally in the Landauer
formula we can ignore the contribution from slow electrons
provided that the energy interval of integration 
is much larger than  $|M|$.

%********************************************************************
%\subsection*{Current}
%********************************************************************
According to the Landauer approach \cite{Landauer:PS:92}
we have, at zero temperature, 
the following expression for the current.
\begin{eqnarray}
I=\frac{2e}{h} \int_0^{E_F}dE\sum_{n} T_{n}(E)
\;.
\label{eq:current}
\end{eqnarray}
For $T_{n}(E)$ which is the total transmission probability we
have:
%({\em eq:tottransm})
\begin{eqnarray}
  \label{eq:tottransm}
%\lefteqn{
T_{n}(E)
%}\hspace{5mm}
= \sum_m  W_{mn}(E) |t_{mn}(E)|^2 \Theta (E-E_n) 
\;,
\end{eqnarray}
where $W_{mn}(E)$ is the 
barrier transmission probability for an electron in mode $m$
at energy $E+ l_{mn}\hbar\omega$.
Assuming the barrier to be smooth 
we can ignore tunneling and use the following expression:
%({\em eq:barrtrans})
\begin{eqnarray}
  \label{eq:barrtrans}
%\lefteqn{
W_{mn}(E)
%}\hspace{5mm}
=\Theta \left[E+l_{nm}\hbar \omega  -(E_b+E_m(x_b))\right]
\;.
\end{eqnarray}
Here $E_b$ is the potential energy due to the line-gate
and $E_m(x_b)$ is the residual effective
mode potential induced by the transverse confinement
in the widened region.
We consider the case when $E_b$
is adjusted so that
there is no transmission in absence of the FIR field,
and we ignore $E_m(x_b)$.
The part that is reflected from the barrier
need not be accounted for since it
will travel in the
reverse direction until it reaches the reservoir.

%********************************************************************
\section{Pair-coupling approximation}
%********************************************************************
We now restrict our attention
to situations in which:
%
%({\em eq:paircrit})
%
\begin{eqnarray}
\label{eq:paircrit}
\left|(E_{m'}-E_m)-(E_m-E_{n})\right|
\gg |M|
\;,
\end{eqnarray}
for all sets of three different modes: $m'$, $m$ and $n$.
\cite{squarewell}
In this case, only pair coupling is important.
If there are some modes: $m'$, $m$ and $n$ that do not satisfy
the condition in Eq. (\ref{eq:paircrit}),
a pair-coupling approach is invalid for
frequencies that couple these levels.

Two different kinds of coupling can be distinguished
in the pair-coupling approximation.
There is an ``absorption coupling'' 
in which an electron entering 
in mode $n$ at an energy $E$ is coupled strongly
to a higher mode $m$ at a higher energy $E+\hbar\omega$.
There is also an ``emission coupling'' in which the
same electron is coupled
strongly to a lower mode $m$ at a lower energy $E-\hbar\omega$.
These couplings are not active simultaneously
with our restriction in Eq. (\ref{eq:paircrit}) on the
spectrum.

The ``absorption coupling'' and the ``emission coupling'' give
the same eigenvalue equation by suitable substitutions.
For the ``absorption coupling'' we have $l_{nn}=0$ and $l_{mn}=1$,
and we introduce:
%({\em eq:abssubst})
%
\begin{eqnarray}
\bar{K}_{mn}^{ABS}(E)&=&\left(K_{m1}+K_{n0}\right)/2
\nonumber \\
\Delta K_{mn}^{ABS}&=&K_{m1}-K_{n0}
\nonumber \\
\tilde{\cal K}^{ABS}&=&{\cal K}^{ABS}-\bar{K}_{mn}^{ABS}(E)
\;,
\label{eq:abssubst}
\end{eqnarray}
while for the ``emission coupling'' we have
$l_{nn}=0$ and $l_{mn}=-1$ and we introduce:
\begin{eqnarray}
\bar{K}_{mn}^{EM}(E)&=&\left(K_{m(-1)}+K_{n0}\right)/2
\nonumber \\
\Delta K_{mn}^{EM}&=&K_{m(-1)}-K_{n0}
\nonumber \\
\tilde{\cal K}^{EM}&=&{\cal K}^{EM}
-\bar{K}_{mn}^{EM}(E)
\;.
\end{eqnarray}

Introducing $M=M_{nm}=-M_{mn}$
and dropping subscripts and superscripts for now,
we get the following
symmetric and $E$-independent form of Eq. (\ref{eq:smalleigsys}):
\begin{equation}
\left(
\begin{array}{cc}
-\Delta K/2
&
-iM
\\
iM
&
\Delta K/2
\end{array}
\right)
\left(
\begin{array}{c}
C_n
\\
C_m
\end{array}
\right)
=
\tilde{\cal K}
\left(
\begin{array}{c}
C_n
\\
C_m
\end{array}
\right)\;.
\label{eq:matrixequation}
\end{equation}
The eigenvalues are: $\tilde{\cal K}^{(\pm)} = 
\pm \sqrt{(\Delta K/2)^2+M^2}$ and
the orthonormal eigenvectors are:
%({\em eq:egenvekt})
%
\begin{eqnarray}
\left(
\begin{array}{c}
C_n
\\
C_m
\end{array}
\right)^{(+)}
=
\left(
\begin{array}{c}
-ia
\\
b
\end{array}
\right)
\;\;,\;\;
\left(
\begin{array}{c}
C_n
\\
C_m
\end{array}
\right)^{(-)}
=
\left(
\begin{array}{c}
\;\;ib\;
\\
\;\;a\;
\end{array}
\right)
\;,
\label{eq:egenvekt}
\end{eqnarray}
\newpage
\noindent
where:
\begin{eqnarray}
a&=&
\frac{1}{\sqrt{2}}
\;\sqrt{1-\sin \alpha}
\nonumber \\
b&=&
\frac{1}{\sqrt{2}}
\;\sqrt{1+\sin \alpha}
\nonumber \\
\sin\alpha
&=&
\frac{\Delta K/2}{\sqrt{(\Delta K/2)^2+M^2}}
\;.
\end{eqnarray}

Using the eigenvectors in Eq. (\ref{eq:egenvekt}) we find the
expansion coefficients using the matching condition in 
Eq. (\ref{eq:match}):
\begin{eqnarray}
a_n^{(+)}=ia
\;\;,\;\;
a_n^{(-)}=-ib
\;.
\end{eqnarray}
Then we have for the transition probability
in Eq. (\ref{eq:tottransm}):
%({\em eq:Rabiformula})
%
\begin{eqnarray}
|t_{mn}(E)|^2
&=&
\left\{
\begin{array}{ccc}
\gamma_{mn} \sin ^2
\left(
q_{mn}(E)L
\right)
&,&\bar{K}_{mn}(E)\geq0
\\
0\;&,&\bar{K}_{mn}(E)<0
\end{array}
\right.
\,,
\nonumber
\\
\gamma_{mn}
&=&
\frac{M_{mn}^2}{\left(\Delta K_{mn}/2\right)^2+M_{mn}^2}
\;,
\nonumber \\
q_{mn}(E)
&=&
\frac{P_{n,E}^{(+)}-P_{n,E}^{(-)}}{2\hbar}
\approx
\frac{\pi}{\bar{\lambda}}\,\sqrt\frac{\left(\Delta K_{mn}/2\right)^2
+M_{mn}^2}{\bar{K}_{mn}^2(E)}
\;,
\nonumber \\
\bar{\lambda}
&=&
\bar{\lambda}_{mn}(E)
=
\frac{h}{ \sqrt{2m^*\bar{K}_{mn}(E)}}
\;.
\label{eq:Rabiformula} 
\end{eqnarray}

We have already 
assumed a large kinetic energy: $K_{n0}\gg |M|$, in the
incident mode.
Then, if $\bar{K}<0$ we know that the kinetic energy
in the other mode must be negative.
This implies a large kinetic energy mismatch: 
$|\Delta K |\gg |M|$, and thus corresponds to
an off-resonant situation ($\gamma\ll 1 $).
Thus we are allowed to ignore the coupling and
put $|t_{mn}(E)|^2=0$ when $\bar{K}<0$, instead
of trying to find an exact expression.

The approximation for $q_{mn}(E)$ 
in Eq. (\ref{eq:Rabiformula}) is good when
$\left(\Delta K/2\right)^2
+M^2 \ll \bar{K}^2$
However, since $K_{n0}\gg |M|$ we 
find that the approximation for $q_{mn}(E)$ is poor only in
off-resonant situations: $\left(\Delta K/2\right)^2
\gg M^2$.
Therefore we can use this approximation generally.

From Eq. (\ref{eq:Rabiformula}) we see that the population
oscillates between the incident mode $n$ to the other mode $m$
along the channel. This is similar to Rabi oscillation in time
of a two-level system.
The wavelength of oscillation is given by $\pi/q$,
and the resonance strength is given by $\gamma$. 
It is clear from the expression for $\gamma$ that
the pair comes to resonance when: $|\Delta K| < 2M$.

\begin{figure}[htb] \begin{center}\leavevmode
\epsfxsize 0.8 \hsize
\epsfbox{
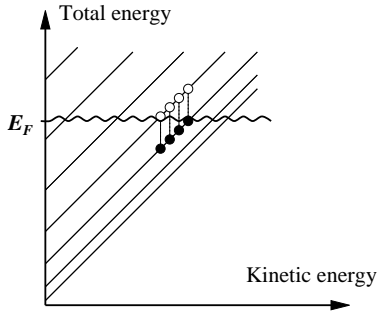
}
\vspace{-0.5ex}
\caption{
\label{fig:pairs}
In the limit of weak coupling
and zero temperature, only pair transitions
close to the Fermi level matters. 
In this case an analytical expression
for the transition amplitudes can be found.
}
\end{center}
\end{figure}

The following relations are easily verified:
\begin{eqnarray}
\bar{K}_{mn}^{ABS}(E)&=&\bar{K}_{nm}^{EM}(E+\hbar\omega)
\nonumber \\
\Delta K_{mn}^{ABS}&=&\Delta K_{nm}^{EM}
\;,
\end{eqnarray}
which leads to:
%
%({\em eq:balance})
\begin{eqnarray}
|t_{mn}^{ABS}(E)|^2&=&|t_{nm}^{EM}(E+\hbar\omega)|^2
\;.
\label{eq:balance}
\end{eqnarray}
Equation (\ref{eq:balance}) states that as the electron
from $|n,E\rangle$ is pumped up there is another electron
from $|m,E+\hbar\omega\rangle$, which is pumped down,
completely canceling the net change in the mode population.
This is provided that both the scattering states are occupied.
The cancelation which is a manifestation of the
orthogonality between scattering states, appears when all
coupled modes are occupied.
For this reason only pumping between states that have
different occupation factors will give a net oscillation
along the channel.
At zero temperature this happens for electrons that
are closer than $\hbar\omega$ to the Fermi level
(see Fig. \ref{fig:pairs}).

We consider the case: $T=0$, $E_b=E_F$ and $E_m(x_b)=0$.
In this case only electrons that have been 
excited into an empty state will
be transmitted over the barrier. 
It is clear that only ``absorption coupling''
can lead to this and we find the following expression for
the current:
%({\em eq:current2})
%
\begin{eqnarray}
I=
\sum_{n=1}^N
\sum_{m>n}^M
I_{mn}
\;,
\label{eq:current2}
\end{eqnarray}
with the partial currents given by:
%({\em eq:current3})
%
\begin{eqnarray}
I_{mn}=
\frac{2e}{h} \int_{E_F-\hbar\omega}^{E_F}
 |t_{mn}^{ABS}(E)|^2 \Theta(E-E_n) dE 
\;,
\label{eq:current3}
\end{eqnarray}
In the sums in Eq. (\ref{eq:current2}),
$N$ is the number of propagating modes and $M$ can be chosen
to fulfill:
$E_{M}-E_N \geq \hbar\omega$.
For a symmetric confining potential there are no transitions
for which $m+n=$ even.
The transition probabilities $|t_{mn}^{ABS}(E)|^2$ are 
found from Eq. 
(\ref{eq:Rabiformula}) where for the ``absorption coupling''
we have from Eqs. (\ref{eq:abssubst}) and 
(\ref{eq:kinenergy})
%({\em eq:absexpl})
%
\begin{eqnarray}
\bar{K}_{mn}(E)&=&E+\frac{\hbar\omega-(E_m+E_n)}{2}
\nonumber \\
\Delta K_{mn}&=&\hbar\omega-(E_m-E_n)
\;,
\label{eq:absexpl}
\end{eqnarray}

From Eqs. (\ref{eq:current3}), (\ref{eq:Rabiformula}) and 
(\ref{eq:absexpl}) it is clear that the 
pair resonances give rise
to peaks in the current for certain frequencies $f_{mn}=(E_m-E_n)/h$.
The width of these peaks is: $\Delta f_{mn}=2M_{mn}/h$.

%
%********************************************************************
\section{Results}
%********************************************************************
We will here consider the case of a square well confining 
potential with impenetrable walls at $y=\pm d/2$. In this
case we have from Eq. (\ref{eq:Matrixelement}) the
following transition elements:
%({\em eq:sqm})
\begin{eqnarray}
  \label{eq:sqm}
M_{mn}
&=&
\frac{4 \hbar e \hat{E}}{3\omega d m^*}
\frac{3mn}{2(m^2-n^2)}
\;
\delta_{m+n,odd}
\nonumber
\end{eqnarray}

\begin{figure}[htb] \begin{center}\leavevmode
\epsfxsize 0.9 \hsize
%\epsfbox{
%/home/mes/tageman/projekt.dir/pbhfet.dir/figures.dir/2Dcurrapprox.eps
%}
\epsfbox{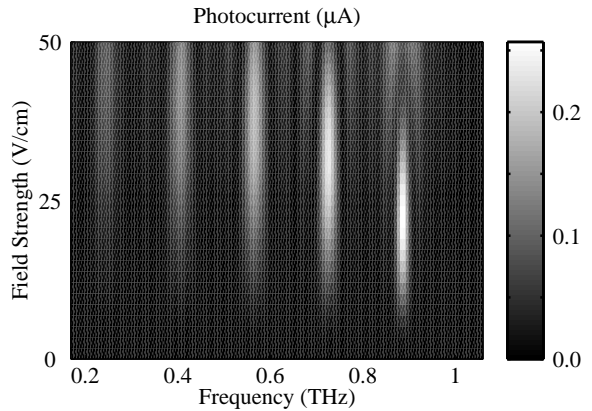
}
\vspace{1.5ex}
\caption{
\label{fig:currplot}
Plot of the current versus both $\nu=\omega/2\pi$ and $\hat{E}$
for a $5\,\mu$m long channel with six propagating modes.
Peaks (light areas) correspond to pair resonances.
The right column gives the scale for the current in microamps. 
}
\end{center}
\end{figure}

First we will demonstrate the frequency and field strength 
dependence of the current found
within the pair-coupling approximation
(see Eq. (\ref{eq:current2})).
In Fig. (\ref{fig:currplot}) we plot the current
for a channel for which $L=5\, \mu$m, $N=6$ and $E_F=14$meV.

For weak FIR fields the pair-resonance peaks are well separated
and the pair coupling approximation is good.
However, in the upper-right part of the figure there is
a slight overlap,
and using the general scattering state 
in Eq. (\ref{eq:scattsol}) 
would lead to a more accurate  result in this case.

%********************************************************************
%\subsection*{Gate voltage dependence}
%********************************************************************

In the situation where the confining potential is
tunable by means of gates,
it is interesting to consider the effect of varying the
gate voltage, $V_g$.
The confining potential can change in different ways 
when $V_g$ is changed.
As an example we will consider what happens if only the
width $d$ is changed, still using a square well potential.
Then we have:
%
%({\em eq:modenvg})
%
\begin{eqnarray}
  \label{eq:modenvg}
E_n=E_F\frac{n^2}{N^2(V_g)}
\;,
\end{eqnarray}
where $N(V_g)$ still is the number of propagating modes. 
In order to make connection to the famous split-gate experiment
by van Wees et. al 
\cite{Wees:PRL:88}, we use: $N(V_g)=20+9V_g$.

\begin{figure}[htb] \begin{center}\leavevmode
\vspace{3ex}
\epsfxsize0.9\hsize
%\epsfbox{/home/mes/amritpal/Projekt/Figures/singleFreq_fast.eps}
\epsfbox{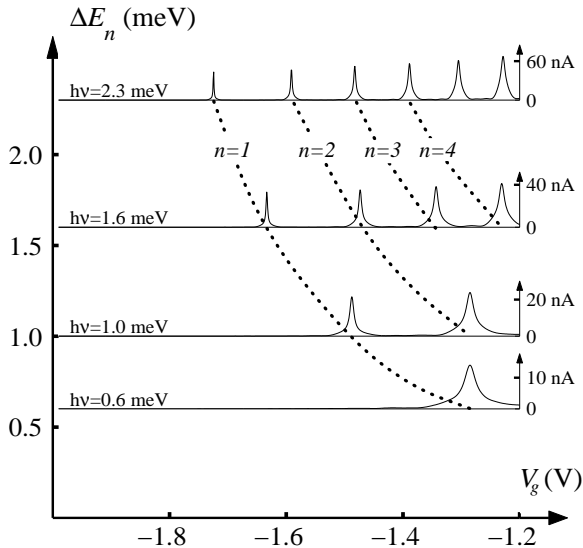}
%\epsfbox{/home/mes/tageman/projekt.dir/pbhfet.dir/figures.dir/singfdekor.eps}
\vspace{1ex}
\caption{
\label{fig:singleFreq}
By sweeping $V_g$ for a set of frequencies and
interpolating between the positions of resonance peaks in the current
one can find out how  the mode energy separations
$\Delta E_{n}= E_{n+1}-E_n$ change with $V_g$.
}
\end{center}
\end{figure}

We can now observe the current peaks 
without sweeping the frequency.
By instead sweeping the gate voltage we change the
mode energy separation $E_m-E_n$ and thus the
resonance frequencies. 
When $E_m-E_n$ matches the
applied frequency a pair will come to resonance.
In Fig. \ref{fig:singleFreq} we
show the result of sweeping $V_g$ for a set of frequencies,
when $\hat{E}=5$ V/cm.
In this case one can verify that the 
condition on the spectrum in Eq. (\ref{eq:paircrit}) is
clearly fulfilled
in the considered frequency range.

Each frequency gives its own set of peaks.
It is clear that for a high frequency it takes a more negative
$V_g$ to separate the mode energies sufficiently.
Therefore the peaks are shifted to the left with
increasing frequency in the diagrams.
By drawing a line that interpolates this shift of the
peaks we get a complete picture of how the mode energy
separations $\Delta E_{n}= E_{n+1}-E_n$ changes with $V_g$. 
By adding these
separations we get the full mode spectrum
relative to the lowest mode, $E_n(V_g)-E_1(V_g)$.

%********************************************************************
\section{Discussion}
%********************************************************************

In order to have a coherent influence on the propagating electrons
the coherence time in the far infrared source must
exceed the passage time of electrons. 
A highly coherent source is typically not widely tunable.
In experiments of the considered kind, fixed
frequency FIR lasers are used.
\cite{Janssen:JPC:94,Arnone:APL:95,Wyss:APL:95}
Therefore it is experimentally advantageous that 
we can reconstruct the mode spectrum
using only fixed frequency sources, and also see how it
changes with $V_g$.
By the interpolation method one can gain information about
all occupied modes for a given $V_g$, and not only 
about the one which is closest to pinch-off.
Theoretical considerations suggest that there is a transition
from a parabolic confinement towards a more square-well like, as
the channel is made wider.
\cite{Laux:SS:88}
Such detailed information is not found by measuring only the
conductance versus $V_g$ since, due to a variation in 
the charge density, the spectrum changes with $V_g$.
\cite{Wees:PRL:88,Wharam:JPC:88}
For channels with a parabolic confining potential,
$V(y)=m^*\omega^2_0y^2/2$, 
magnetotranport experiments can
be used to determine $\omega_0$. 
\cite{Berggren:PRB:88}

From a physical point of view our system is interesting
in that it enables a directed acceleration of electrons 
caused by a high frequency field, via a two step process.
First energy is absorbed in standing wave excitations
across the channel. In this process a strong influence is
possible since both energy and momentum along the channel
is conserved at resonance.
Then the energy stored in the transverse direction is 
released in the widening and adiabatically converted 
into kinetic energy along the channel.

A nonzero temperature will destroy our predictions in two 
ways. 
First, the coherence length will be reduced because of
enhanced phonon scattering. In experiments clear 
quantized conductance in a 5 $\mu$m channel has been demonstrated
at 1.3 K.
\cite{Tarucha:JJAP:95}
Since such temperatures are  used also in FIR-transport 
experiments this seems within reach.
\cite{Janssen:JPC:94}
The second degradation comes from thermal smearing. As a criterion
we can take that the temperature must be much smaller then the 
inter mode separation, which is typically of the order 20 K.

We have ignored collective effects.
In absorption experiments on wire arrays one
finds a depolarization effect.
\cite{Ando:RMP:82}
As a result the mode spectrum found from absorption
spectroscopy can differ from that found from
magnetoresistance oscillations.
\cite{Brinkop:PRB:88}
This is not immediately generalized to a
non-homogeneous system.
A difference is that we consider 
coherent transport and that the inhomogeneity acts as
a boundary condition for propagating electrons.
We can not rule out a significant influence
due to depolarization, but leave it
for future investigations.

Another collective effect
is the static spatial modulation of the potential
along the channel because
of changes in the mode population. However such an
influence is suppressed in a multi mode channel since only
a fraction of the electrons are resonantly coupled and
electrons moving with different velocities give depopulation 
at different locations along the channel.
In addition most charge comes from slow electrons, and
for these strong cancellation can be expected because of a
strong energy dependence.

In conclusion we consider the influence of a
FIR field on coherent transport in a multi mode
channel.
We focus on a long channel in which resonance conditions
prevail over a considerable distance
in order to achieve a high
sensitivity to an applied FIR field.
We derive both a general scattering state and 
an approximation valid in the limit of weak fields,
where pair resonance will dominate.
In this limit the current will peak when $\hbar\omega=E_m-E_n$.
We also propose a way to analyze the spectrum, using fixed
frequencies only.

%******************************************************************
\section{Acknowledgment}
%******************************************************************
We acknowledge financial support from the EU (MEL ARI Research
project 22953 - CHARGE) and from
the Swedish Research Council
for Engineering Sciences (TFR). Ola Tageman is supported by
Ericsson Microwave Systems AB.
We are grateful to L. Y. Gorelik
for his valuable advise.

%******************************************************************
%\section{Appendix}
%******************************************************************

%In a short channel the resonance peaks will not 
%dominate over the non resonant background.
%In order to detect the first complete inversion of population
%we must use a channel of length 
%$L\gg 2\lambda\cdot|K/\Delta S_n|$.
%However in the limit $LM/\lambda K \ll 1$ we find the less
%restrictive criterion: $L^2\gg( 4\lambda\cdot|K/\Delta S_n|)^2$.

%esonances develops on a length scale 
%$\lambda_R=\lambda \cdot K/M$,
%while non resonant coupling develops on 
%length scale $\lambda_N=\lambda \cdot 2K/\Delta K$, which
%can be much shorter especially in the weak coupling limit.
%In the worst case the population change due to non resonant
%coupling is fully developed and thus of a strength given by:
%$\gamma\approx (4M/\Delta S)^2$, where we again look at the
%situation midway between peaks.

\vspace{-0.5cm}
%******************************************************************
%\section{References}
%******************************************************************


\begin{thebibliography}{9}

\vspace{-1.5cm}
\bibitem{Wyss:APL:93}
R. A. Wyss, C. C. Eugster, J. A. del Alamo, and Q. Hu, 
Appl. Phys. Lett. {\bf 63}, 1522 (1993).

\bibitem{Janssen:JPC:94}
T. J. Janssen, J. C. Maan, J. Singleton, N. K. Patel, M. Pepper, 
J Phys C.: Condens. Matter, {\bf 6}, 
L163 (1994).

\bibitem{Wyss:APL:95}
R. A. Wyss, C. C. Eugster, J. A. del Alamo, and Qing. Hu,
M. J. Rooks, and M. R. Melloch, 
Appl. Phys. Lett. {\bf 66}, 1144 (1995).

\bibitem{Arnone:APL:95}
D. D. Arnone, J. E. F. Frost, C. G. Smith, D. A. Ritchie, 
G. A. C. Jones,
R. J. Butcher, and M. Pepper,
Appl. Phys. Lett. {\bf 66},
3149 (1995).

\bibitem{Wees:PRL:88}
B.~J.~van Wees, H.~van Houten, C.~W.~J. Beenakker,
J.~G.~Williamson, L.~P.~Kouwenhoven, D.~van der Mare,
and C.~T.~Foxon, Phys. Rev. Lett. {\bf 60}, 848 (1988).

\bibitem{Wharam:JPC:88}
D. A. Wharam, T. J. Thornton, R.~Newbury, M.~Pepper,
H.~Ahmed, J.~E.~F.~Frost, D.~G.~Hasko, D.~C.~Peacock,
D.~A.~Ritchie, and G.~A.~C.~Jones, J. Phys. C{\bf 21},
L209 (1988).

%\bibitem{Hu:APL:93}
%Qing. Hu,
%Appl. Phys. Lett. {\bf 62},
%837 (1993).

%\bibitem{Tien:PR:63}
%P. K. Tien, and J. P. Gordon,
%Phys. Rev. {\bf 129}, 647 (1963).

%\bibitem{Hu:SST:96}
%Qing Hu, S. Verghese, R. A. Wyss, Th. Schäpers,
%J. del Alamo, S. Feng, K. Yakubo, M. J. Rooks, M. R. Melloch,
%and A Förster
%Semicond. Sci. Technol. {\bf 11}, 1888 (1996).

%\bibitem{Feng:PRB:93}
%S.Feng and Q. Hu
%,Phys. Rev. B {\bf 48},
%5354 (1993).

\bibitem{Hekking:PRB:91}
F. Hekking, and Yu. V. Nazarov,
Phys. Rev. B {\bf 44},
9110 (1991).

\bibitem{Gorelik:PRL:94}
L. Y. Gorelik, Anna. Grincwajg, 
V. Z. Kleiner, R. I. Shekhter, and M. Jonson,
Phys. Rev. Lett. {\bf 73}, 2260 (1994).

\bibitem{Grincwajg:PRB:95}
Anna Grincwajg, L. Y. Gorelik, V. Z. Kleiner, and R. I. Shekhter,
Phys. Rev. B {\bf 52}, 12168  (1995).

\bibitem{Tarucha:JJAP:95}
T. Honda, S. Tarucha, T. Saku, and Y. Tokura,
Jpn. J. Appl. Phys. {\bf 34}, L72 (1995).

\bibitem{Tageman:JAP:97}
Ola Tageman, L.Y. Gorelik, R. I. Shekhter, and M. Jonson,
J. Appl. Phys. {\bf 81}, 285 (1997).
L. Y. Gorelik, M. Jonson, R. I. Shekhter, and O. Tageman,
Proceedings of the NATO Advanced Study Institute on 
Frontiers in Nanoscale Science of 
Micron/Submicron Devices, Kiev August 1995
(NATO ASI Series E: Applied Sciences, Vol. 328).

\bibitem{Tageman:JAP:98}
Ola Tageman, and L.Y. Gorelik,
J. Appl. Phys. {\bf 83}, 1513 (1998).

\bibitem{Glazman:JETP:88}
L. I. Glazman, G.B. Lesovik, D.E. Khmel'nitskii, and R.I. Shekhter
JETP Lett. {\bf 48}, 238 (1988).

\bibitem{Glazman:PRB:90}
L. I. Glazman and M. Jonson, Phys. Rev. B{\bf 41},
10686 (1990).

\bibitem{Landauer:PS:92}
R. Landauer, Physica Scripta T{\bf 42}, 110 (1992).

\bibitem{squarewell}
If none of the modes: $m'$, $m$ and $n$ enter the channel,
the condition in Eq. (\ref{eq:paircrit}) does not
have to be fulfilled for these modes.
For a hard-wall potential of width $d$ one can prove:
$\left|(E_{m'}-E_m)-(E_m-E_{n})\right| \geq 
\hbar^2\pi^2/2m^*d^2$.

\bibitem{Laux:SS:88}
S. E.Laux, D. J. Frank, and Frank Stern,
Surf. Sci. {\bf 196}, 101 (1988).

\bibitem{Berggren:PRB:88}
K.-F. Berggren, G. Roos, and H. van Houten
Phys. Rev. B {\bf 37}, 10118 (1988).

\bibitem{Ando:RMP:82}
T. Ando, A. B. Fowler, and F. Stern,
Rev. Mod. Phys. {\bf 54}, 437 (1982).

\bibitem{Brinkop:PRB:88}
F. Brinkop, W. Hansen, and J. P. Kotthaus
Phys. Rev B {\bf 37}, 6547 (1988).




\end{thebibliography}
\end{document}